\documentclass{article}
\usepackage[english]{babel}
\usepackage[letterpaper,top=2cm,bottom=2cm,left=3cm,right=3cm,marginparwidth=1.75cm]{geometry}
\usepackage{amsthm,amssymb,amsmath}  
\usepackage{mathtools}
\usepackage{graphicx}
\usepackage[colorlinks=true, allcolors=blue]{hyperref}
\usepackage[capitalize]{cleveref}
\usepackage{xspace}
\usepackage{thmtools}
\usepackage{thm-restate}
\usepackage{url}
\usepackage{caption}
\usepackage{subcaption}
\usepackage{authblk}
\usepackage{colortbl}
\usepackage{color}

\def\dd{\mathinner{.\,.}}
\newcommand{\cO}{\mathcal{O}}

\newcommand{\Occ}{\textsf{Occ}}
\newcommand{\PrefSuf}{\textsf{PrefSuf}}
\newcommand{\ST}{\textsf{ST}\xspace}

\definecolor{mygray}{rgb}{0.863,0.863,0.863}
\newcommand{\mybox}[2]
	{\noindent \\
		\colorbox{mygray}{
			\fbox{\begin{minipage}{.95\textwidth}{\bf #1 \\[2mm]}#2\end{minipage}
			}\\[2mm]
		}
	}

\newtheorem{theorem}{Theorem}

\newtheorem{observation}{Observation}
\newtheorem{example}{Example}

\title{Text Indexing and Pattern Matching with Ephemeral Edits}
\author[1,2]{Solon P.\ Pissis}
\affil[1]{CWI, Amsterdam, The Netherlands}
\affil[2]{Vrije Universiteit, Amsterdam, The Netherlands}

\begin{document}

\maketitle

\begin{abstract} A sequence $e_0,e_1,\ldots$ of edit operations in a string $T$ is called \emph{ephemeral} if operation $e_i$ constructing string $T^i$, for all $i=2k$ with $k\in\mathbb{N}$, is reverted by operation $e_{i+1}$ that reconstructs $T$. Such a sequence arises when processing a stream of \emph{independent} edits or testing \emph{hypothetical} edits.

We introduce \emph{text indexing with ephemeral substring edits}, a new version of text indexing. The goal is to design a data structure over a given text that supports subsequent pattern matching queries with ephemeral \emph{substring} insertions, deletions, or substitutions in the text; we require insertions and substitutions to be of constant length. In particular, we preprocess a text $T=T[0\dd n)$ over an integer alphabet $\Sigma=[0,\sigma)$ with $\sigma=n^{\cO(1)}$ in $\cO(n)$ time. Then, we preprocess any arbitrary pattern $P=P[0\dd m)$ given online in $\cO(m\log\log m)$ time and $\cO(m)$ space and support any ephemeral sequence of edit operations in $T$. Before reverting the $i$th operation, we report all \textsf{Occ} occurrences of $P$ in $T^i$ in $\cO(\log\log n + \textsf{Occ})$ time.

We also introduce \emph{pattern matching with ephemeral substring edits}, a new version of pattern matching. The goal is to design a data structure over two given strings that supports pattern matching with ephemeral \emph{substring} insertions, deletions, or substitutions in one of the two strings; we require insertions and substitutions to be of constant length. In particular, we preprocess two strings $T$ and $P$, each of length at most $n$, over an integer alphabet $\Sigma=[0,\sigma)$ with $\sigma=n^{\cO(1)}$ in $\cO(n)$ time. Then, we support any ephemeral sequence of edit operations in $T$. Before reverting the $i$th operation, we report all \textsf{Occ} occurrences of $P$ in $T^i$ in the optimal $\cO(\textsf{Occ})$ time. Along our way to this result, we also give an optimal solution for pattern matching with ephemeral \emph{block deletions}. 

The utility of these results is underscored by the following key aspects:
\begin{itemize}
    \item Our algorithms are founded on a model highly motivated by real-world applications; e.g., in pangenomics, one often wants to process local variants originating from \emph{different individuals}.
    \item As anticipated, our algorithms \emph{outperform} their fully dynamic counterparts [Gawrychowski et al., SODA 2018; Amir et al., ACM Trans. Algorithms 2007] within this model.
    \item In stark contrast to their fully dynamic counterparts, our algorithms are \emph{simple} to implement, leveraging only textbook algorithms or data structures with mature implementations.
\end{itemize}
\end{abstract}

\newpage

\section{Introduction}

Pattern matching and text indexing are the two flagship problems of combinatorial pattern matching.
In \emph{pattern matching}, we are given a pattern $P=P[0\dd m)$ of length $m$ and a text $T=T[0\dd n)$ of length $n$, and we are asked to find all occurrences of $P$ in $T$. This can be solved in the optimal $\cO(n)$ time using the classical Knuth-Morris-Pratt algorithm~\cite{DBLP:journals/siamcomp/KnuthMP77}. In \emph{text indexing}, we are asked to preprocess a text $T=T[0\dd n)$ into a data structure that supports subsequent pattern matching queries: given a pattern $P=P[0\dd m)$, we report (or count) all occurrences of $P$ in $T$. Reporting all $\textsf{Occ}$ occurrences of $P$ in $T$ can be done in the optimal $\cO(m + \textsf{Occ})$ time using suffix trees, which can be constructed in $\cO(n)$ time~\cite{DBLP:journals/jacm/Farach-ColtonFM00}. 

A central question of theoretical and practical interest is:
\begin{center}
  \emph{What happens when text $T$ is subjected to edit operations?} 
\end{center}

\subparagraph{Motivation.} In a typical setting, one already has at hand a long reference string as the text $T$ (e.g., a reference genome, a baseline legal document or a configuration file), which will remain mostly static, except for a few local edits. These local edits are often \emph{independent} (e.g., genomic variants coming from individuals or client-specific variations); or \emph{hypothetical} (e.g., they are not intended to be permanent, but rather are applied temporarily to test a specific scenario in a scientific simulation or to analyze the impact of a potential alteration). More concretely, this setting is highly relevant in computational pangenomics~\cite{DBLP:journals/bib/Consortium18,Tettelin2020Pangenome,10.1093/bib/bbae588}, where $T$ models an \emph{existing} reference sequence, and the sequence of independent edit operations represents a collection of short local variants originating from \emph{different individuals}.\footnote{The Variant Call Format (\url{https://en.wikipedia.org/wiki/Variant_Call_Format}) stores this type of data. In fact, many international consortia (e.g., \url{https://alpaca-itn.eu/} and \url{https://www.pangenome.eu/}) deal with such data.} The fact that local variants originate from different individuals suggests that once a variant is processed, the corresponding edit should be reverted before processing the next one. Indeed, when querying a pattern against one individual's genome (which is a variant of the reference), \underline{we do not want that variant to persist} when moving on to query against another individual's genome, which will have its own distinct short local variant from the reference.

In response, we introduce new versions of text indexing and pattern matching with ephemeral edits. 
A sequence $e_0,e_1,\ldots$ of edit operations in a string $T$ is called \emph{ephemeral} if operation $e_i$ constructing string $T^i$, for all $i=2k$ with $k\in\mathbb{N}$, is reverted by operation $e_{i+1}$ that reconstructs $T$. Since for any odd $i$, $T^{i}=T$, when the context is clear, we may simply use $T^i$ to refer to a case where $i$ is even.

\subparagraph{Our Contributions.} We introduce a new interesting version of text indexing that we call \emph{text indexing with ephemeral substring edits}. The goal is to design a data structure over a given text that supports subsequent pattern matching queries with ephemeral \emph{substring} insertions, deletions, or substitutions in the text; we require insertions and substitutions to be short (of constant length), but deletions can be arbitrarily long. First, we preprocess a text $T$. Then, we are asked to preprocess any arbitrary pattern $P$ given online, and subsequently allow any ephemeral sequence of edit operations in $T$. Before reverting the $i$th operation, we must report all occurrences of $P$ in $T^i$. We show that after an $\cO(n)$-time preprocessing of text $T$ of length $n$, we can preprocess any arbitrary pattern $P$ of length $m$ in $\cO(m \log\log m)$ time and $\cO(m)$ space to then support pattern matching queries after any edit in $\cO(\log\log n + \textsf{Occ})$ worst-case time. Note that $P$ is not static; we can preprocess any subsequent pattern in $\cO(m \log\log m)$ time and $\cO(m)$ space (i.e., independently of $n$). Further note that our query time is essentially independent of $m$. More formally, we show the following result.  

\begin{theorem}\label{the:dynamic}
    Given a text $T$ of length $n$ over an integer alphabet $\Sigma=[0,\sigma)$ with $\sigma=n^{\cO(1)}$, we can preprocess it in $\cO(n)$ time and space to support the following:
    \begin{description}
        \item For any pattern $P$ of length $m$ and any constant $\epsilon\geq 1$ given online, we can preprocess $P$ in $\cO(m\log\log m)$ time and $\cO(m)$ space to support the following sequence of queries:
    \begin{enumerate}
        \item[(1)] Insert string $S$ of length $|S|\leq \epsilon$ after position $p$ in $T$, for any $-1\leq p<n$, or delete fragment $T[q\dd p]$, for any $0\leq q\leq p<n$, or substitute fragment $T[p\dd p+|S|-1]$ by string $S$ of length $|S|\leq \epsilon$, for any $0\leq p\leq n-|S|$, to obtain text $T'$ in $\cO(1)$ worst-case time.
        \item[(2)] Report all $\textsf{Occ}$ occurrences of $P$ in $T'$ in $\cO(\log\log n + \textsf{Occ})$ worst-case time.
        \item[(3)] Revert the edit operation in (1) to obtain $T$ in $\cO(1)$ worst-case time.
    \end{enumerate}
    \end{description}
\end{theorem}

Our technique underlying \Cref{the:dynamic} consists in a novel $\cO(m\log\log m)$-time preprocessing of $P$ into a collection of trees of $\cO(m)$ total size. This preprocessing allows us to deploy \emph{predecessor search}~\cite{DBLP:journals/csur/NavarroR20} and optimal \emph{prefix-suffix queries}~\cite{DBLP:conf/sosa/Pissis25,DBLP:journals/talg/Gawrychowski13} to process each edit in near-optimal time. With \Cref{the:dynamic}, we can support efficient pattern matching queries while processing ephemeral sequences of edits in $T$.

We also introduce a new version of pattern matching that we call \emph{pattern matching with ephemeral block deletions}. The goal is to design a data structure over two given strings $T$ and $P$, each of length at most $n$, that supports pattern matching with ephemeral block deletions in $T$. First, we preprocess $T$ and $P$ in $\cO(n)$ time. Then, we support any ephemeral sequence of block deletions in $T$. Before reverting the $i$th operation, we report all occurrences of $P$ in $T^i$ in the optimal $\cO(\textsf{Occ})$ worst-case time. More formally, we show the following result.  

\begin{theorem}\label{the:pm-del}
    Given a text $T$ and a pattern $P$, each of length at most $n$, over an integer alphabet $\Sigma=[0,\sigma)$ with $\sigma=n^{\cO(1)}$, we can preprocess $T$ and $P$ in $\cO(n)$ time and space to support the following sequence of queries:
    \begin{enumerate}
        \item[(1)] Delete $T[q\dd p]$, for any $0\leq q\leq p<|T|$, 
        to obtain text $T'$ in $\cO(1)$ worst-case time.
        \item[(2)] Report all $\textsf{Occ}$ occurrences of $P$ in $T'$ in $\cO(\textsf{Occ})$ worst-case time.
        \item[(3)] Revert the edit operation in (1) to obtain $T$ in $\cO(1)$ worst-case time.
    \end{enumerate}
\end{theorem}

Our data structure underlying \Cref{the:pm-del} relies on \emph{suffix trees}~\cite{DBLP:journals/jacm/Farach-ColtonFM00} and optimal prefix-suffix queries.

Finally, we introduce another version of pattern matching that we call \emph{pattern matching with ephemeral substring edits}. The goal is to design a data structure over two given strings $T$ and $P$, each of length at most $n$, that supports pattern matching with ephemeral \emph{substring} insertions, deletions, or substitutions in $T$; we require insertions and substitutions to be short (of constant length), but deletions can be arbitrarily long, as indeed we rely on \Cref{the:pm-del}. First, we preprocess $T$ and $P$ in $\cO(n)$ time. Then, we support any ephemeral sequence of edit operations in $T$. Before reverting the $i$th operation, we report all occurrences of $P$ in $T^i$ in the optimal $\cO(\textsf{Occ})$ worst-case time.  More formally, we show the following result.  

\begin{theorem}\label{the:pm-edit}
    Given a text $T$ and a pattern $P$, each of length at most $n$, over an integer alphabet $\Sigma=[0,\sigma)$ with $\sigma=n^{\cO(1)}$, we can preprocess $T$ and $P$ in $\cO(n)$ time and space to support the following sequence of queries:
    \begin{enumerate}
        \item[(1)] Insert string $S$ of length $|S|=\cO(1)$ after position $p$ in $T$, for any $-1\leq p<|T|$, or delete fragment $T[q\dd p]$, for any $0\leq q\leq p<|T|$, or substitute fragment $T[p\dd p+|S|-1]$ by string $S$ of length $|S|=\cO(1)$, for any $0\leq p\leq |T|-|S|$, to obtain text $T'$ in $\cO(1)$ worst-case time.
        \item[(2)] Report all $\textsf{Occ}$ occurrences of $P$ in $T'$ in $\cO(\textsf{Occ})$ worst-case time.
        \item[(3)] Revert the edit operation in (1) to obtain $T$ in $\cO(1)$ worst-case time.
    \end{enumerate}
\end{theorem}

Our data structure underlying \Cref{the:pm-edit} relies on a somewhat novel combination of suffix trees and \emph{string matching automata}~\cite{DBLP:books/daglib/0020103} as well as on optimal prefix-suffix queries.

The algorithms underlying \Cref{the:dynamic,the:pm-del,the:pm-edit} are also \emph{simple} to implement: they utilize only textbook algorithms or data structures with mature implementations; e.g., predecessor search~\cite{DBLP:conf/wea/Dinklage0H21}, suffix trees~\cite{DBLP:conf/wea/GogBMP14}, string matching automata~\cite{DBLP:books/daglib/0020103}, and range data structures~\cite{DBLP:conf/wea/GogBMP14}. 

\subparagraph{Related Work.} A natural question that arises is: 

\begin{center}
    \emph{Why cannot one use the existing dynamic text indexing algorithms to solve the above problems?}

\end{center}
To the best of our knowledge, the existing dynamic text indexing algorithms have seen limited to no adoption by practitioners. One reason for this is certainly their extreme intricacy, as exemplified in~\cite{DBLP:conf/esa/LiptakM024}. Another related reason is that, while a fully dynamic model is theoretically appealing, it often proves overly powerful for the above problems. 
Let us review the main works on the topic before we provide more concrete arguments to answer the above question.
The problem of \emph{dynamic text indexing} was introduced by Gu, Farach, and Beigel in 1994~\cite{DBLP:conf/soda/GuFB94}. The problem asks to design a data structure (a dynamic index) that supports updates (insertions or deletions) to a text $T$ and subsequent pattern matching queries. The solution of~\cite{DBLP:conf/soda/GuFB94} (see also~\cite{DBLP:conf/isaac/AmirB20} for an amortized analysis) achieves $\cO(\log n)$-time updates to $T$ and $\cO(m + \textsf{Occ}\log i + i\log m)$-time queries, where $i$ is the number of edit operations executed so far and $m$ is the length of the pattern. Ferragina~\cite{DBLP:journals/jal/Ferragina97} later showed how to support insertions and deletions of text blocks with the query time being roughly proportional to the number of updates made so far. Soon after that, Ferragina and Grossi~\cite{DBLP:journals/jal/FerraginaG99} improved this to support updates in $\cO(\sqrt{n})$ time and queries in the optimal $\cO(m + \textsf{Occ})$ time. Sahinalp and Vishkin~\cite{DBLP:conf/focs/SahinalpV96} showed the first polylogarithmic data structure that supports $\cO(\log^3 n)$-time updates and $\cO(m + \textsf{Occ} + \log n)$-time queries. The update time was later improved by Alstrup, Brodal, and Rauhe~\cite{DBLP:conf/soda/AlstrupBR00} to $\cO( \log^2 n \log \log n \log^{\star} n)$ at the expense of slightly slower queries with $\cO(m + \textsf{Occ} + \log n \log \log n)$ time. Finally, this was improved by Gawrychowski et al.~\cite{DBLP:journals/corr/GawrychowskiKKL15,DBLP:conf/soda/GawrychowskiKKL18} to $\cO(\log^2 n)$-time updates and $\cO(m + \textsf{Occ})$-time queries. 
More recently, Kempa and Kociumaka~\cite{DBLP:conf/stoc/KempaK22}
proposed a dynamic version of suffix arrays~\cite{DBLP:journals/siamcomp/ManberM93}---which is strictly 
harder than dynamic text indexing---supporting $\cO(\log^4 n)$-time updates.

\mybox{Why is \Cref{the:dynamic} useful?}{Say that we want to perform $k$ \emph{independent} or \emph{hypothetical} edits on an existing text $T$ of length $n$ and report all occurrences of a pattern $P$ of length $m$ after each edit. If we use the state-of-the-art algorithm for fully dynamic text indexing~\cite{DBLP:journals/corr/GawrychowskiKKL15} as a black box, the total time will be $\cO((n+k)\log^2(n+k) + km + \textsf{Output})$. Using \Cref{the:dynamic} instead, the total time will be $\cO(n + m\log\log m + k\log\log n + \textsf{Output})$. Note that our improvement is substantial: even if we ignore the polylogarithmic factors, we replace the $km$ term by $k$.} 

Amir et al.~\cite{DBLP:journals/talg/AmirLLS07} have also considered a restricted version of \emph{dynamic pattern matching}, where the pattern $P$ is \emph{static} (i.e., it is always the same) and only letter substitutions are supported in $T$. The preprocessing time of their data structure is $\cO(n \log \log m+ m \sqrt{\log m})$ and the space used is $\cO(n + m \sqrt{\log m})$. After each text update, the algorithm deletes all previous occurrences of the pattern that no longer match, and reports all new occurrences of the pattern in the text in $\cO(\log \log m)$ time.

\mybox{Why is \Cref{the:pm-edit} useful?}{Say that we want to perform $k$ \emph{independent} or \emph{hypothetical} substitutions on an existing text $T$ of length $n$ and report all occurrences of a pattern $P$ of length $m$ after each edit. If we use the algorithm for a dynamic text and a static pattern~\cite{DBLP:journals/talg/AmirLLS07}, the total time and space will be $\cO(n\log\log m + m\sqrt{\log m} + k\log\log m + \textsf{Output})$ and $\cO(n + m\sqrt{\log m})$, respectively. Using \Cref{the:pm-edit} instead, the total time and space will be $\cO(n + k + \textsf{Output})$ and $\cO(n)$, respectively. Again, our improvement is substantial (both in time and in space).} 

Although ephemeral edits have already been used for many other string problems, such as \emph{longest common substrings}~\cite{DBLP:conf/spire/AmirCIPR17,DBLP:journals/algorithmica/AmirCPR20}, \emph{palindrome substrings}~\cite{DBLP:conf/cpm/FunakoshiNIBT18,DBLP:conf/cpm/FunakoshiNIBT19,DBLP:conf/spire/FunakoshiM21,DBLP:journals/tcs/FunakoshiNIBT21}, \emph{Lyndon substrings}~\cite{DBLP:conf/cpm/UrabeNIBT18}, and \emph{substring covers}~\cite{DBLP:conf/cpm/MitaniMSH24}, such a model has (to the best of our knowledge) not been defined for text indexing or pattern matching. We were thus also motivated to close this (surprising) gap in theory. 

\subparagraph{Paper Organization.} In \Cref{sec:prel} we provide the necessary definitions and notation, as well as a few important tools that we use throughout. In \Cref{sec:indexing}, we present our solution to text indexing with ephemeral substring edits (\Cref{the:dynamic}). In \Cref{sec:app-pm}, we present our solution to pattern matching with ephemeral block deletions (\Cref{the:pm-del}). In \Cref{sec:app-pm-edit}, we present our solution to pattern matching with ephemeral substring edits (\Cref{the:pm-edit}). We conclude this paper in \Cref{sec:fin} with some future proposals.

\section{Preliminaries}\label{sec:prel}

\subparagraph{Strings.} We consider finite strings over an integer \emph{alphabet} $\Sigma=[0,\sigma)$. 
The elements of $\Sigma$ are called \emph{letters}.
For a string $T = T[0]\cdots T[n-1]$ over alphabet $\Sigma$,
its \emph{length} is $|T| = n$. 
For any $0 \leq i \leq j < n$, the string $T[i] \cdots T[j]$ is called a \emph{substring} of $T$. 
By $T[i\dd j]$ we denote its occurrence at (starting) position $i$, and we call it a \emph{fragment} of $T$. 
When $i = 0$, this fragment is called a
\emph{prefix} (we may denote it by $T[\dd j]$). When $j = n-1$, this fragment is called a \emph{suffix} (we may denote it by $T[i\dd ]$). We may use $T[i\dd j)$ to denote the fragment $T[i\dd j-1]$ of $T$ or $T(i\dd j]$ to denote the fragment $T[i+1\dd j]$ of $T$.

\subparagraph{String Indexes.} Given a set $\mathcal{S}$ of strings, a \emph{trie} $\textsf{T}(\mathcal{S})$ is a rooted tree whose nodes represent the prefixes of strings in $\mathcal{S}$~\cite{DBLP:books/daglib/0020103}. The prefix corresponding to node $u$ is denoted by $\textsf{str}(u)$; and the node $u$ is called the \emph{locus} of $\textsf{str}(u)$. Sometimes we are interested in maximizing this prefix for an arbitrary string $S$. We say that the locus of an arbitrary string $S$ in $\textsf{T}(\mathcal{S})$ is the node $u$ such that $\textsf{str}(u)$ is a prefix of $S$ and $|\textsf{str}(u)|$ is maximized. The parent-child relationship in $\textsf{T}(\mathcal{S})$ is defined as follows: the root node is the locus of the empty string $\varepsilon$; and the parent $u$ of another node $v$ is the locus of $\textsf{str}(v)$ without the last letter. This letter is the label of the edge $(u,v)$. The order on $\Sigma$ induces an order on the edges outgoing from any node of $\textsf{T}(\mathcal{S})$. A node $u$ is called \emph{branching} if it has at least two children and \emph{terminal} if $\textsf{str}(u) \in \mathcal{S}$. 

A \emph{compacted trie} is obtained from $\textsf{T}(\mathcal{S})$ by dissolving all nodes except the root, the branching nodes, and the terminal nodes. The dissolved nodes are called \emph{implicit} while the preserved nodes are called \emph{explicit}. The compacted trie takes $\cO(|\mathcal{S}|)$ space provided that the edge labels are stored as pointers to fragments of strings in $\mathcal{S}$. Given the lexicographic order on $\mathcal{S}$ along with the lengths of the longest common prefixes between any two consecutive (in this order) elements of $\mathcal{S}$, one can compute the compacted trie in $\cO(|\mathcal{S}|)$ time~\cite{DBLP:conf/cpm/KasaiLAAP01}.

The \emph{suffix tree} of a string $T$, denoted by $\textsf{ST}(T)$, is precisely the compacted trie of the set of all suffixes of $T$. Every terminal node in $\textsf{ST}(T)$ is labeled by the starting position 
of its corresponding suffix in $T$.
Suffix trees usually come with suffix links by construction: a \emph{suffix link}
for a branching node $u$ of $\textsf{ST}(T)$, denoted by $\textsf{slink}(u)$, is the node $v$ such that $\textsf{str}(v)$ is the longest proper suffix of $\textsf{str}(u)$, i.e., if $\textsf{str}(u) = T[i \dd j]$ then $\textsf{str}(v)=T (i \dd j]$. The \emph{suffix array} of $T$, denoted by $\textsf{SA}(T)$, is the array of the suffixes of $T$ in lexicographic order. It is often useful to define the \emph{inverse suffix array} as $\textsf{iSA}[\textsf{SA}[i]] = i$, for all $i\in[0,n)$. The data structures $\textsf{ST}(T)$, $\textsf{SA}(T)$, and $\textsf{iSA}(T)$ can be constructed in $\cO(n)$ time for any string $T$ of length $n$ over an integer alphabet $\Sigma=[0,\sigma)$ with $\sigma=n^{\cO(1)}$~\cite{DBLP:journals/jacm/Farach-ColtonFM00}. 
The edges of $\textsf{ST}(T)$ can be accessed in $\cO(1)$ time if stored using perfect hashing~\cite{DBLP:journals/jacm/BenderCFKT23}. 

\subparagraph{Prefix-Suffix Queries.} Let $T$ be a string of length $n$ over an integer alphabet $\Sigma=[0,\sigma)$. We would like to preprocess $T$ to support the following type of queries, known as \emph{prefix-suffix queries}~\cite{DBLP:conf/soda/GuFB94}: for any $i,j \in [0,n)$ return all $\Occ$ occurrences of $T$ in $T[0\dd i]T[j\dd n)$. This type of query, which we denote here by $\PrefSuf(i,j)$, can be simply and efficiently implemented due to the following recent result.

\begin{theorem}[\cite{DBLP:conf/sosa/Pissis25}]\label{the:PSQ}
    For any string $T$ of length $n$ over an integer alphabet $\Sigma=[0,\sigma)$ with $\sigma=n^{\cO(1)}$,
    we can answer any $\PrefSuf(i,j)$ query in $\cO(1)$ time after an $\cO(n/\log_\sigma n)$-time preprocessing. The data structure size is $\cO(n/\log_\sigma n)$ and the output is given as a compact representation (a single arithmetic progression) of $\cO(1)$ size.
\end{theorem}

In fact, the data structure of \Cref{the:PSQ}~\cite{DBLP:conf/sosa/Pissis25} can be constructed in $\cO(n)$ time, for any alphabet $\Sigma$, if the \emph{longest common extension} (LCE) data structure of~\cite{DBLP:conf/stoc/KempaK19} is replaced by the classical Z-algorithm~\cite{DBLP:books/cu/Gusfield1997}. This is possible because one of the two indices for the \emph{longest common prefix} queries asked is $0$ and one of the two indices for \emph{longest common suffix} queries is $n-1$.

\subparagraph{Predecessor Search.} Assume we have a set $X$ of $n$ keys from a universe $U=[0,u)$ with a total order. In the \emph{predecessor} problem, we are given a query element $q\in U$, and we are asked to find the maximum $p\in X$ such that $p \leq q$ (the predecessor of $q$)~\cite{DBLP:journals/csur/NavarroR20}. When $X$ is \emph{static}, and its elements are given \emph{sorted}, the \emph{$y$-fast trie} data structure~\cite{DBLP:journals/ipl/Willard83} can be constructed in $|X|=\cO(n)$ time and space, and it supports predecessor queries in $\cO(\log\log u)$ time; see~\cite{DBLP:journals/csur/NavarroR20} for more details.
Thus, when $u = n$, the query time becomes $\cO(\log\log n)$.

\section{Text Indexing with Ephemeral Substring Edits}\label{sec:indexing}

In this section, we prove \Cref{the:dynamic}.
We design a data structure that supports pattern matching queries with ephemeral substring insertions, deletions, or substitutions. First, we preprocess the text $T=T[0\dd n)$ over alphabet $\Sigma=[0,\sigma)$ with $\sigma=n^{\cO(1)}$. Then, we preprocess any arbitrary pattern $P=P[0\dd m)$ given online and support any ephemeral sequence of edit operations in $T$. Before reverting the $i$th operation, we report all occurrences of $P$ in $T^i$ in $\cO(\log\log n + \textsf{Occ})$ worst-case time. In \Cref{subsec:text}, we present the preprocessing of the text, which occurs one time.
In \Cref{subsec:pattern}, we present the preprocessing of the pattern, which occurs one time per arriving pattern. In \Cref{subsec:edit}, we present the algorithm for processing the edit operations. Let us first start with describing the high-level idea.

\subparagraph{High-Level Idea.} Suppose an edit operation $e_i$ occurs at some position in $T$. Our goal is to efficiently locate the \emph{longest prefix} of $P$ ending right before the edit and the \emph{longest suffix} of $P$ spanning the edit. It is then sufficient to trigger one prefix-suffix query. For locating the occurrences of $P$ not spanning the edit (they are before or after the edit, and so they are not affected by it), we use a standard approach from computational geometry. We focus on finding the longest such suffix of $P$ (finding the longest such prefix of $P$ is symmetric). 

When a pattern $P$ arrives, we find the union of occurrences of the suffixes of $P$ in $T$ using the suffix tree of $T$. Although these occurrences can be $\Omega(m)$, we can encode them using at most $m$ intervals on $T$. These intervals are not pairwise disjoint because a suffix of $P$ can be a prefix of another suffix of $P$. We can, however, make them disjoint and still have $\cO(m)$ pairwise disjoint intervals on $T$. The invariant is that every occurrence from $T$ is associated only with the \emph{longest suffix} of $P$. Since the intervals are now disjoint, we can insert them in a predecessor data structure. Hence, each edit takes $\cO(\log\log n + \textsf{Occ})$ time.

\subsection{Preprocessing the Text}\label{subsec:text}

\begin{figure}[t]
    \centering
    \includegraphics[width=1\linewidth]{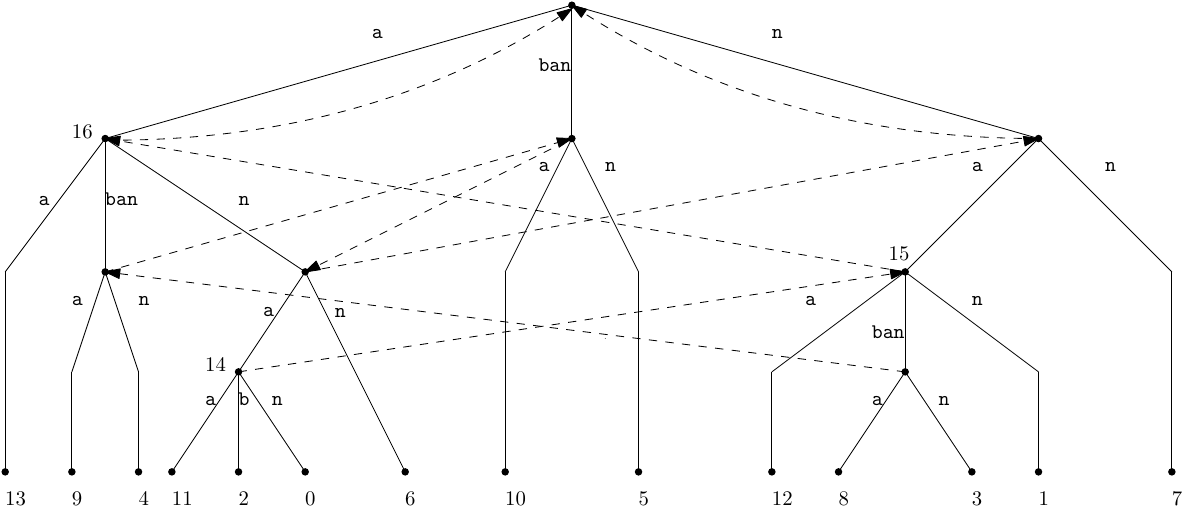}
    \caption{The suffix tree \textsf{ST} of $T=T[0\dd 17)=\texttt{ananabannabanaana}$ with suffix links (dotted). The label of edges leading to leaf nodes is truncated after the first letter to avoid cluttering the figure. The suffix array \textsf{SA} of $T$ is $[16,13,9,4,14,11,2,0,6,10,5,15,12,8,3,1,7]$ and is inferred from \textsf{ST} using an in-order bottom-up traversal. The node labeled 16, which represents suffix $T[16\dd 17)$, stores also the \textsf{SA} interval $[0,8]$. The node spelling string $\texttt{ban}$ from the root stores the \textsf{SA} interval $[9,10]$.}
    \label{fig:ST}
\end{figure}

The preprocessing of text $T$ is standard.    
We construct the suffix tree (with suffix links) $\textsf{ST}(T)$, the suffix array $\textsf{SA}(T)$, and the inverse suffix array $\textsf{iSA}(T)$. This takes $\cO(n)$ time and space~\cite{DBLP:journals/jacm/Farach-ColtonFM00}. We also store, at each node of $\textsf{ST}(T)$, the corresponding $\textsf{SA}(T)$ interval -- this can be achieved in $\cO(n)$ time using a bottom-up traversal on $\textsf{ST}(T)$; inspect \Cref{fig:ST} for an example. We also construct a data structure to answer any range maximum query (RMQ) on $\textsf{SA}(T)$ and a data structure for answering any range minimum query (RmQ) on $\textsf{SA}(T)$. The preprocessing time and space is $\cO(n)$~\cite{DBLP:conf/latin/BenderF00}. This concludes the preprocessing of $T$.

\subsection{Preprocessing the Pattern}\label{subsec:pattern}

The preprocessing of pattern $P$ is more involved.
When $P$ arrives, we construct the data structure of \Cref{the:PSQ} for prefix-suffix queries over $P$. This preprocessing takes $\cO(m)$ time and space.
We also spell $P$ in $\textsf{ST}(T)$ by means of suffix links to find the locus of the longest substring starting at every position of $P$ and occurring in $T$. 
In particular, for every \emph{suffix} $P[i\dd m)$ of $P$ occurring in $T$, we store the suffix array interval $[s,e]_i$ labeled by $i$: $P[i\dd m)$ occurs at positions $\textsf{SA}[s],\ldots,\textsf{SA}[e]$ of $T$.
This can be done in $\cO(m)$ time (provided that we are given $\textsf{ST}(T)$) using the classical \emph{matching statistics} algorithm~\cite{DBLP:books/cu/Gusfield1997}.

\begin{figure}
     \centering
     \begin{subfigure}[b]{0.45\textwidth}
         \centering
         \includegraphics[width=0.7\textwidth]{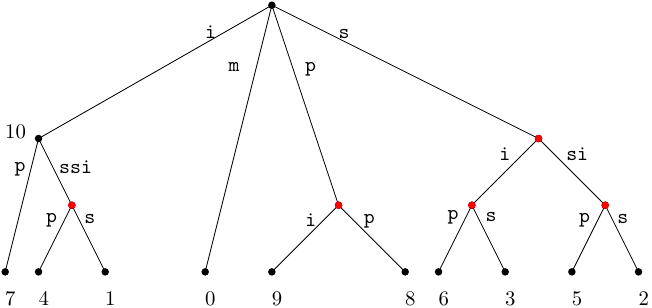}
         \subcaption{$\textsf{ST}(P)$; the nodes colored red are the ones which are going to be dissolved in $\textsf{TREE}(P)$.}
     \end{subfigure}
     \hfill
     \begin{subfigure}[b]{0.45\textwidth}
         \centering
         \includegraphics[width=0.7\textwidth]{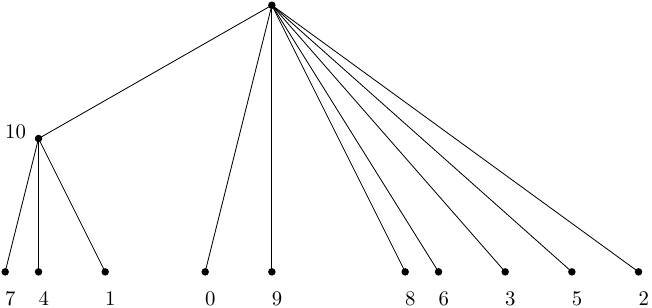}
         \subcaption{$\textsf{TREE}(P)$; the nodes colored red in $\textsf{ST}(P)$ are dissolved here.}
     \end{subfigure}
        \caption{$\textsf{ST}(P)$ and $\textsf{TREE}(P)$ for $P=P[0\dd 11)=\texttt{mississippi}$. The label of edges leading to leaf nodes in $\textsf{ST}(P)$ is truncated after the first letter to avoid cluttering the figure.
        The labels are omitted in $\textsf{TREE}(P)$.}
        \label{fig:treeP}
\end{figure}

We next sort the letters of $P$ in $\cO(m \log\log m)$ time and $\cO(m)$ space~\cite{DBLP:journals/jal/Han04}. (Note that this sorting is required only when $\sigma$ is relatively large; i.e., $\sigma=m^{\cO(1)}$ does not hold.)
Each letter of $P$ is then replaced by its rank; in particular, each letter is now an integer from $[0,m)$. We can thus construct the suffix tree $\ST(P)$ of $P$ in $\cO(m)$ time and space. Given $\ST(P)$, we derive the following rooted tree, which we denote by $\textsf{TREE}(P)$; inspect \Cref{fig:treeP} for an example. 
The tree $\textsf{TREE}(P)$ is on exactly $m+1$ nodes: the root node and one node per suffix of $P$. We start from $\ST(P)$ and dissolve every branching node $u$ that does not represent a suffix of $P$ (along with its incoming edges) and connect every child $v$ of $u$ directly with the parent $\textsf{parent}(u)$ of $u$ via the edge $(\textsf{parent}(u),v)$ labeled by the concatenation of the labels of the edges $(\textsf{parent}(u),u)$ and $(u,v)$. Thus every root-to-node path represents a single suffix of $P$. We have an internal (possibly branching) node if and only if it spells a suffix of $P$ that is a prefix of another suffix (possibly many) of $P$. The edge labels are encoded by intervals over $[0,m)$ and so the size of $\textsf{TREE}(P)$ is $\cO(m)$ and it takes $\cO(m)$ time to construct it using a traversal on $\ST(P)$. This preprocessing takes $\cO(m\log\log m)$ time (or $\cO(m)$ time if we do not sort the letters of $P$) and $\cO(m)$ space.

\begin{figure}
     \centering
     \begin{subfigure}[b]{0.45\textwidth}
         \centering
         \includegraphics[width=0.7\textwidth]{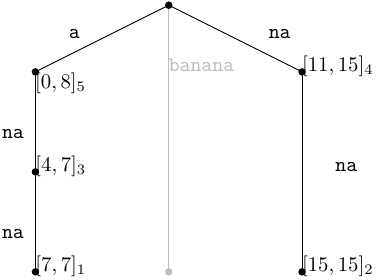}
         \subcaption{$\textsf{TREE}(P)$ with the intervals from the \textsf{SA} of $T$.}
         \label{fig:tree1}
     \end{subfigure}
     \hfill
     \begin{subfigure}[b]{0.45\textwidth}
         \centering
         \includegraphics[width=0.7\textwidth]{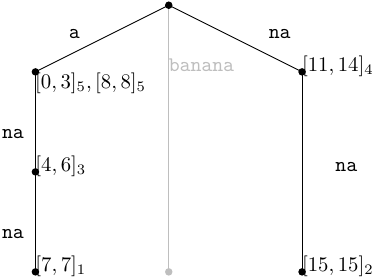}
         \subcaption{$\textsf{TREE}(P)$ after making the intervals disjoint.}
         \label{fig:tree2}
     \end{subfigure}
        \caption{$\textsf{TREE}(P)$ for $P=\texttt{banana}$ with the intervals from the \textsf{SA} of $T=\texttt{ananabannabanaana}$. Note that the suffix $\texttt{banana}$ of $P$ has no interval stored because it does not occur in $T$.}
        \label{fig:tree}
\end{figure}

Using the loci information obtained from $\textsf{ST}(T)$ (via the matching statistics algorithm), each node of $\textsf{TREE}(P)$ representing the suffix $P[i\dd m)$ is associated with the computed interval $[s,e]_i$. Note that, by construction, the interval associated to every node is contained in the intervals associated to its ancestors. We then decompose the elements (indices) of node intervals (i.e., the union of occurrences of suffixes of $P$ in $T$) into a new collection of intervals such that $\textsf{iSA}[j]\in [s',e']_i$ if and only if $P[i\dd m)$ is the \emph{longest suffix} of $P$ occurring as a prefix of $T[j\dd n)$; inspect \Cref{fig:tree} for an example. This can be done for all node intervals using a traversal on $\textsf{TREE}(P)$. The total number of all new intervals obtained from  $\textsf{TREE}(P)$ is still in $\cO(m)$ because the number of new intervals is linear in the number of nodes and edges in the tree. By construction, these newly created intervals are \emph{pairwise disjoint}. Moreover, by construction, these intervals are \emph{sorted} because both $\ST(T)$ and $\ST(P)$ have been constructed using the same order: the one induced by the same alphabet $\Sigma$. For every $[s',e']_i$ in the collection of intervals, we insert $s'$ into a static predecessor data structure (in particular, a static $y$-fast trie~\cite{DBLP:journals/ipl/Willard83}) with satellite values $e'$ and $i$. This preprocessing takes $\cO(m)$ time and space because the $s'$ values to be inserted are sorted.

\begin{example}
Let $T=\texttt{ananabannabanaana}$ and $P=\texttt{banana}$.
We have the following:
\begin{itemize}
    \item suffix \texttt{anana} with interval $[7,7]_1$ in \textsf{SA} is split to $[7,7]_1$
    \item suffix \texttt{ana} with interval $[4,7]_3$ in \textsf{SA} is split to $[4,6]_3$
    \item suffix \texttt{a} with interval $[0,8]_5$ in \textsf{SA} is split to $[0,3]_5$ and $[8,8]_5$ 
    \item suffix \texttt{banana} has no interval stored as it does not occur in $T$
    \item suffix \texttt{nana} with interval $[15,15]_2$ in \textsf{SA} is split to $[15,15]_2$
    \item suffix \texttt{na} with interval $[11,15]_4$ in \textsf{SA} is split to $[11,14]_4$.
\end{itemize}
\end{example}

\begin{figure}[ht]
    \centering
    \includegraphics[width=.7\linewidth]{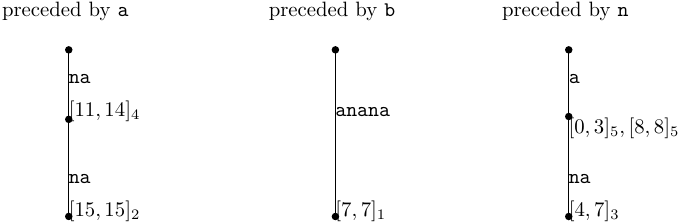}
    \caption{$\textsf{TREE}_c(P)$ obtained from $\textsf{TREE}(P)$ for $P=\texttt{banana}$ and $c\in\{\texttt{a},\texttt{b},\texttt{n}\}$ with the intervals from the \textsf{SA} of $T=\texttt{ananabannabanaana}$ made disjoint.}
    \label{fig:treec}
\end{figure}

We also construct (at most) $|\Sigma|$ rooted trees, denoted by $\textsf{TREE}_c(P)$, for all $c\in\Sigma$ that occur in $P$; inspect \Cref{fig:treec} for an example. Specifically, $\textsf{TREE}_c(P)$ is the tree induced by $\textsf{TREE}(P)$
such that every node in $\textsf{TREE}_c(P)$ represents a suffix whose preceding letter is $c$. 
All trees $\textsf{TREE}_c(P)$ can be constructed from the suffix tree $\ST(P)$ of $P$ in $\cO(m)$ total time. 
In one bottom-up traversal, we add the suffixes $P[i\dd m)$, with $c=P[i-1]$, in a group labeled $c$.
The suffixes in each group are added in lexicographic order.
We then construct one compacted trie for each group of suffixes in linear time~\cite{DBLP:conf/cpm/KasaiLAAP01}. For each compacted trie, we dissolve the branching nodes that do not represent a suffix of $P$ (similar to $\textsf{TREE}(P)$) to obtain $\textsf{TREE}_c(P)$.
We then decompose the node intervals in $\textsf{TREE}_c(P)$ into a new collection of intervals such that $\textsf{iSA}[j]\in [s',e']_i$ if and only if $P[i\dd m)$ is preceded by $c$ and $P[i\dd m)$ is the longest suffix of $P$ occurring as a prefix of $T[j\dd n)$. This can be done in a way similar to $\textsf{TREE}(P)$ using a traversal on $\textsf{TREE}_c(P)$. The total size of all trees $\textsf{TREE}_c(P)$ is in $\cO(m)$ because if a letter $c'$ from $\Sigma$ does not occur in $P$ then no tree $\textsf{TREE}_{c'}(P)$ is created.
The total number of all intervals obtained from all trees $\textsf{TREE}_c(P)$ is in $\cO(m)$. Due to how we construct them, the intervals from a single tree are pairwise disjoint and sorted.
For every $[s',e']_i$ in the collection of intervals for $\textsf{TREE}_c(P)$, we insert $s'$ into a separate static $y$-fast trie labeled by $c$ with satellite values $e'$ and $i$. This preprocessing takes $\cO(m)$ time and space.

\begin{example}
Let $T=\texttt{ananabannabanaana}$ and $P=\texttt{banana}$. We have the following:

\begin{itemize}
\item Preceded by $\texttt{a}$:
\begin{itemize}
    \item suffix \texttt{na} with interval $[11,15]_4$ in \textsf{SA} is split to $[11,14]_4$
    \item suffix \texttt{nana} with interval $[15,15]_2$ in \textsf{SA} is split to $[15,15]_2$
\end{itemize}
\item Preceded by $\texttt{b}$:
\begin{itemize}
    \item suffix \texttt{anana} with interval $[7,7]_1$ in \textsf{SA} is split to $[7,7]_1$
\end{itemize} 
\item Preceded by $\texttt{n}$:
\begin{itemize}
    \item suffix \texttt{a} with interval $[0,8]_5$ in \textsf{SA} is split to $[0,3]_5$ and $[8,8]_5$ 
    \item suffix \texttt{ana} with interval $[4,7]_3$ in \textsf{SA} is split to $[4,7]_3$.
\end{itemize}
  
\end{itemize}
\end{example}

\begin{figure}[ht]
    \centering
    \includegraphics[width=.7\linewidth]{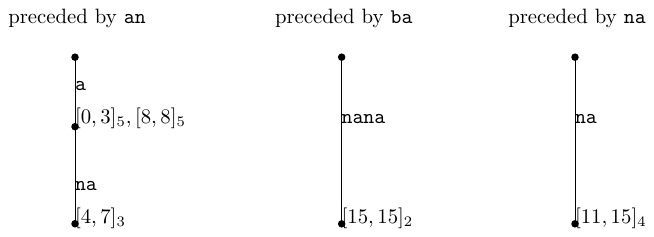}
    \caption{$\textsf{TREE}_S(P)$ obtained from $\textsf{TREE}(P)$ for $P=\texttt{banana}$ and $S\in\{\texttt{an},\texttt{ba},\texttt{na}\}$ with the intervals from the \textsf{SA} of $T=\texttt{ananabannabanaana}$ made disjoint.}
    \label{fig:treeS}
\end{figure}

Let $\epsilon\geq 1$ be a constant given together with pattern $P$.
It should be clear that the construction of $\textsf{TREE}_c(P)$ can be generalized to $\textsf{TREE}_S(P)$, for any string $S$ of length $|S|\geq 1$. In particular, we construct the trees $\textsf{TREE}_S(P)$,
for all $S$ with $|S|\in[1,\lfloor \epsilon \rfloor]$ in $\cO(\epsilon m)=\cO(m)$ time and space. This is possible because, for any fixed length $|S|$, there are $\cO(m)$ suffixes of $P$ to be grouped; inspect \Cref{fig:treeS} for an example.

Thus the preprocessing of pattern $P$ takes $\cO(m\log\log m)$ time and $\cO(m)$ space. Note that the preprocessing takes $\cO(m)$ time and space when $\sigma=m^{\cO(1)}$.

\subsection{Edit Operation}\label{subsec:edit}

Let us start by classifying the occurrences of $P$ in $T^{i}$. We will use this classification to argue about the correctness of our algorithm. We have the following occurrence types:
\begin{description}
    \item[Type 1]: The starting and ending positions are both before the edit
    \item[Type 2]: The starting and ending positions are both after the edit
    \item[Type 3]: The starting position is before the edit and the ending position is within the edit
    \item[Type 4]: The starting position is within the edit and the ending position is after the edit
    \item[Type 5]: The starting position is before the edit and the ending position is after the edit
    \item[Type 6]: The starting and ending positions are both within the edit.
\end{description}

\begin{figure}[ht]
     \centering
     \begin{subfigure}[b]{0.45\textwidth}
         \centering
         \includegraphics[width=0.9\textwidth]{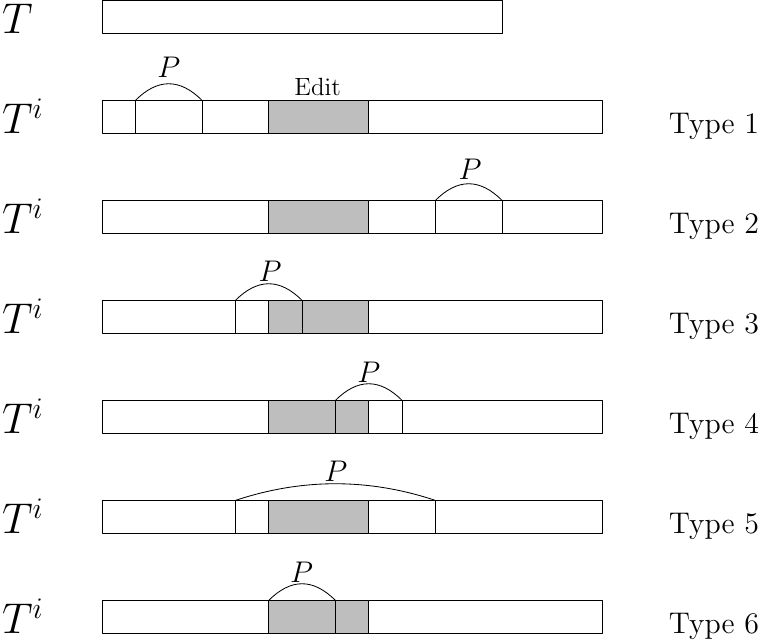}
         \subcaption{Type of occurrences after one insertion.}
         \label{fig:ins}
     \end{subfigure}
     \hfill
     \begin{subfigure}[b]{0.45\textwidth}
         \centering
         \includegraphics[width=0.9\textwidth]{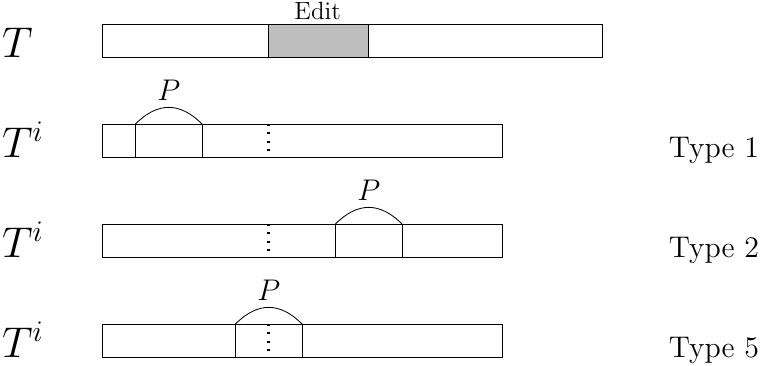}
         \subcaption{Type of occurrences after one deletion.}
         \label{fig:del}
     \end{subfigure}
        \caption{Type of occurrences of a pattern $P$ in text $T^{i}$ after one edit in $T$.}
        \label{fig:types}
\end{figure}

For insertions or substitutions, all the above occurrence types are relevant, whereas for deletions only Types 1, 2, and 5 are relevant. Inspect \Cref{fig:types} for an illustration. It is readily verifiable that these are all possible occurrence types.

For starters, let us assume that $\epsilon=1$ and suppose that an edit operation happens at position $p$ of $T$. We first consider the case of a letter deletion; i.e., we are asked to delete letter $T[p]$.
We find the occurrences of $P$ occurring in $T^{i}=T[0\dd p)T[p+1\dd n)$ as follows. We first find the occurrences of $P$ in $T[0\dd p)$ (\textbf{Type 1}) and in $T[p+1\dd n)$ (\textbf{Type 2}) by using the standard procedure for 3-sided
queries: we retrieve the \textsf{SA} interval of $P$ in $T$ (we have this from the preprocessing part of $P$) 
and use RmQ and RMQ queries recursively on the interval until the answer of a query exceeds the bounds $[0,p-m]$ (for RmQs) and $[p+1,n-m]$ (for RMQs). 
This takes $\cO(1+\textsf{Occ})$ time because every query takes $\cO(1)$ time~\cite{DBLP:conf/latin/BenderF00} and we stop after two unsuccessful queries (one RmQ and one RMQ) or if the intervals are empty.

We still need to find the occurrences spanning $T[\dd p-1]\cdot T[p+1\dd]$ (\textbf{Type 5}). We make the following observation. If we had the longest prefix of $P$ occurring as a suffix of $T[0\dd p)$ and the longest suffix of $P$ occurring as a prefix of $T[p+1\dd n)$, then we could use a single prefix-suffix query, which takes $\cO(1)$ time (\Cref{the:PSQ}), to report all the occurrences of $P$ spanning $T[\dd p-1]\cdot T[p+1 \dd]$. For locating the longest such suffix of $P$ (finding the prefix is symmetric by reversing $T$ and $P$ and so we omit the details), we observe the following.

\begin{observation}[Deletion]\label{obs:del}
    Suppose that we had the suffix tree of $T\#P$, $\#\notin \Sigma$, and
    that we had marked the nodes of the suffix tree representing the suffixes of $P$. The longest suffix of $P$ starting at position $p+1$ of $T$ is the deepest marked ancestor of the node representing $T[p+1\dd n)$ in the suffix tree of $T\#P$.
\end{observation}

Instead of \emph{marked ancestors}~\cite{DBLP:conf/focs/AlstrupHR98}, we use the predecessor data structure over the intervals. Recall that $\textsf{iSA}[p+1]\in [s,e]_i$ if and only if $P[i\dd m)$ is the longest suffix of $P$ that occurs as a prefix of $T[p+1\dd n)$. We find the predecessor $s$ of $\textsf{iSA}[p+1]$ and check if $\textsf{iSA}[p+1]\in[s,e]_i$. 

\subparagraph{Deletion of $T[p]$.} We find the longest suffix of $P$ occurring as a prefix of $T[p+1\dd n)$ as follows. When $p+1$ is given, we find the lexicographic rank (lex-rank) $r=\textsf{iSA}[p+1]$ of $T[p+1\dd n)$ among all suffixes of $T$. We then search for $r$ in the predecessor structure to infer $[s,e]_i$ such that $r\in[s,e]_i$. 
We first locate $s$ as the predecessor of $r$ and then check if $r\in[s,e]_i$.
This gives us the longest suffix $P[i\dd m)$ of $P$ occurring as a prefix of $T[p+1\dd n)$.
Recall that to find the longest prefix of $P$ occurring as a suffix of $T[0\dd p-1]$ we do the symmetric work on the reverse of $P$ and $T$. After finding the longest prefix of $P$ occurring as a suffix of $T[0\dd p-1]$ and the longest suffix of $P$ occurring as a prefix of $T[p+1\dd n)$, we trigger one prefix-suffix query to report all occurrences spanning the deletion. The total time for processing the deletion is $\cO(1)$ and for reporting is $\cO(\log\log n + \textsf{Occ})$, where the $\log\log n$ term is due to predecessor search. The algorithm is correct by \Cref{obs:del}.

\begin{example}[Deletion]
Let $T=\texttt{ananabannabanaana}$ and $P=\texttt{banana}$. Suppose we delete letter $T[p]$ for $p=13$, $T[13]=\texttt{a}$ to obtain $T^{i}=\texttt{ananabanna\underline{banana}}$. The lex-rank of $T[p+1\dd n)=T[14\dd 17)$ is $r=4$. The predecessor of $r=5$ is $s=4$ with $e=6$ and $i=3$ corresponding to suffix $\texttt{ana}$. Similarly, we find the longest prefix of $P$ occurring as a suffix of $T[0\dd p-1]=T[0\dd 12]$ which is $\texttt{ban}$. Thus $P$ occurs at position $10$ of $T^{i}$.
\end{example}

We now consider the case of inserting the letter $c$ after $T[p]$.
We need to find the occurrences of $P$ spanning $T[\dd p]cT[p+1\dd]$ (\textbf{Type 5}) using a prefix-suffix query.
However, for insertions, things are a bit more complicated: we need to locate the longest suffix of $P$ that occurs as a prefix of $cT[p+1\dd n)$.
We make the following observation for locating the longest such suffix of $P$ (finding the prefix is analogous to deletion, and so we omit it).

\begin{observation}[Insertion]\label{obs:ins}
    Suppose that we had the suffix tree of $T\#P$, $\#\notin \Sigma$,
    and that we had marked the nodes of the suffix tree representing the suffixes of $P$. The longest suffix $P[i-1]P[i\dd m)$ of $P$ starting at position $p+1$ of $T^{i}$, for some $i$, is such that the string $P[i\dd m)$ is represented by the deepest marked ancestor of the node representing $T[p+1\dd n)$ in the suffix tree of $T\#P$ and $P[i-1]=T^{i}[p+1]$.
\end{observation}

Again, instead of marked ancestors~\cite{DBLP:conf/focs/AlstrupHR98}, we use the predecessor data structure over the intervals of group $c=P[i-1]$. Recall that $\textsf{iSA}[p+1]\in [s,e]_i$ if and only if $P[i\dd m)$ is preceded by letter $c$ and $P[i\dd m)$ is the longest suffix of $P$ occurring as a prefix of $T[p+1\dd n)$. So we can find the predecessor $s$ of $\textsf{iSA}[p+1]$ and then check if $\textsf{iSA}[p+1]\in[s,e]_i$.

\subparagraph{Insertion of Letter $c$ After $T[p]$.} We find the longest suffix of $P$ occurring as a prefix of $cT[p+1\dd n)$ as follows. When $p+1$ is given, we find the lex-rank $r=\textsf{iSA}[p+1]$ of $T[p+1\dd n)$ among all suffixes of $T$. We then search for $r$ in the predecessor structure for group $c$ to infer $[s,e]_i$, such that $r\in[s,e]_i$, if it exists. This gives us the longest suffix $cP[i\dd m)$ of $P$ occurring as a prefix of $cT[p+1\dd n)$.
Recall that to find the longest prefix of $P$ that is a suffix of $T[0\dd p]$ we do the same work as for the deletion operation. After finding the longest prefix of $P$ that is a suffix of $T[0\dd p]$ and the longest suffix of $P$ occurring as a prefix of $cT[p+1\dd n)$, we trigger one prefix-suffix query. The total time to process the insertion is $\cO(1)$ and for reporting is $\cO(\log\log n + \textsf{Occ})$, where the $\log\log n$ term is due to predecessor search. The algorithm is correct by \Cref{obs:ins}.

\begin{example}[Insertion]
Let $T=\texttt{ananabannabanaana}$ and $P=\texttt{banana}$. We insert letter $c=\texttt{a}$ after $T[p]$ for $p=7$ to obtain $T^{i}=\texttt{anana\underline{ban\textcolor{red}{a}na}banaana}$. The lex-rank of $T[p+1\dd n)=T[8\dd 17)$ is $r=13$. The predecessor of $r=13$ in group $c=\texttt{a}$ is $s=11$ with $e=14$ and $i=4$ corresponding to suffix $\texttt{na}$. Similarly, we find the longest prefix of $P$ that is a suffix of $T[0\dd p]=T[0\dd 7]$ which is $\texttt{ban}$. Thus $P$ occurs at position $5$ of $T^{i}$.
We insert letter $c=\texttt{b}$ as the first letter (i.e., $p=-1$) to obtain $T^{i}=\texttt{\underline{\textcolor{red}{b}anana}bannabanaana}$. We have $p+1=0$ and the lex-rank of $T[p+1\dd n)=T[0\dd 17)$ is $r=7$.
The predecessor of $r=7$ in group $c=\texttt{b}$ is $s=7$ with $i=1$ which corresponds to suffix $\texttt{anana}$. Thus $P$ occurs at position $0$ of $T^{i}$.
\end{example}

Next, we explain the modifications required when $\epsilon>1$.
This is when Types 3, 4, and 6 become relevant for insertions or substitutions.
For deletions, no modification is required.

\subparagraph{Deletion of Fragment $T[q\dd p]$.} We have that $T^{i}=T[0\dd q)T[p+1\dd n)$. For the occurrences of $P$ in $T[0\dd q)$ (\textbf{Type 1}) we use RmQs and for the occurrences of $P$ in $T[p+1\dd n)$ (\textbf{Type 2})
we use RMQs similar to $\epsilon=1$. For the remaining occurrences of $P$ (\textbf{Type 5}), we find the longest suffix of $P$ which is a prefix of $T[p+1\dd n)$ and the longest prefix of $P$ that is a suffix of $T[0\dd q)$ again similar to $\epsilon=1$. The total time for processing the deletion is $\cO(1)$ and for reporting is $\cO(\log\log n + \textsf{Occ})$.

\subparagraph{Insertion of String $S$ After $T[p]$.} We have that $T^{i}=T[0\dd p]\cdot S \cdot T[p+1\dd n)$. For the occurrences of $P$ in $T[0\dd p]$ (\textbf{Type 1}) we use RmQs and for the occurrences of $P$ in $T[p+1\dd n)$ (\textbf{Type 2}) we use RMQs similar to $\epsilon=1$.
For the occurrences of $P$ with a starting position in $[0, p]$ and an ending position after the edit (\textbf{Type 5}), we find the lex-rank $r=\textsf{iSA}[p+1]$ of $T[p+1\dd n)$ in the predecessor structure coming from $\textsf{TREE}_S(P)$.
This gives us the longest suffix of $P$ occurring as a prefix of $S \cdot T[p+1\dd n)$.
We also find the longest prefix of $P$ that is a suffix of $T[0\dd p]$, just like for $\epsilon=1$, and trigger one prefix-suffix query. We are not done yet, as we may have an occurrence of $P$ with a starting position in $[p+1,p+|S|]$ and an ending position in $[p+|S|+1,|T^{i}|)$ (\textbf{Type 4}).
We find the lex-rank $r=\textsf{iSA}[p+1]$ of $T[p+1\dd n)$ in the predecessor structure derived from $\textsf{TREE}(P)$. This gives us the longest suffix of $P$ occurring as a prefix of $T[p+1\dd n)$. We also find the longest prefix of $P$ that is a suffix of $S$ naively in $\cO(\epsilon^2)=\cO(1)$ time, and trigger another prefix-suffix query.
(\textbf{Type 3} is symmetric to \textbf{Type 4}.)
Further note that \textbf{Type 6} occurrences can be found in $\epsilon=\cO(1)$ time using any linear-time pattern matching algorithm~\cite{DBLP:journals/siamcomp/KnuthMP77}.
The total time to process the insertion is $\cO(1)$ and for reporting is $\cO(\log\log n + \textsf{Occ})$.

\begin{example}[Insertion of Substring]
Let $T=\texttt{ananabannabanaana}$ and $P=\texttt{banana}$. We insert substring $S=\texttt{na}$ after $T[p]$ for $p=11$ to obtain $T^{i}=\texttt{ananabanna\underline{ba\textcolor{red}{na}na}ana}$. The lex-rank of $T[p+1\dd n)=T[12\dd 17)$ is $r=12$. The predecessor of $r=12$ in group $S=\texttt{na}$ is $s=11$ with $e=14$ and $i=4$ corresponding to suffix $\texttt{na}$. Similarly, we find the longest prefix of $P$ that is a suffix of $T[0\dd p]=T[0\dd 11]$ which is $\texttt{ba}$. Thus $P$ occurs at position $10$ of $T^{i}$.
\end{example}

Finally, note that substitutions work the same as insertions, but with a different offset: the only change is that when fragment $T[p\dd p+|S|-1]$ is substituted by string $S$, we find the lex-rank $r=\textsf{iSA}[p+|S|]$ of $T[p+|S|\dd n)$ in the predecessor structure derived from $\textsf{TREE}_S(P)$; and then we find the lex-rank $r=\textsf{iSA}[p+|S|]$ of $T[p+|S|\dd n)$ in the predecessor structure coming from $\textsf{TREE}(P)$. After reporting the occurrences of $P$ in $T^{i}$, we do nothing (to revert the edit); i.e., we proceed to the next edit or preprocess a new pattern. We have arrived at \Cref{the:dynamic}.

\section{Pattern Matching with Ephemeral Block Deletions}\label{sec:app-pm}

In this section, we prove \Cref{the:pm-del}.
We design a data structure for pattern matching with ephemeral block deletions. First, we preprocess two strings $T$ and $P$, each of length at most $n$, over alphabet $\Sigma=[0,\sigma)$ with $\sigma=n^{\cO(1)}$. Then, we allow any ephemeral sequence of arbitrarily-long block deletions in $T$. Before reverting the $i$th operation, we report all \textsf{Occ} occurrences of $P$ in $T^i$ in the optimal $\cO(\textsf{Occ})$ worst-case time. This improves upon \Cref{the:dynamic} for this setting.

In \Cref{subsec:preprocess}, we show the preprocessing of $T$ and $P$. In \Cref{subsec:del}, we show the algorithm for processing the block deletions.

\subsection{Preprocessing}\label{subsec:preprocess}

We start by constructing the data structure of \Cref{the:PSQ} for prefix-suffix queries over $P$.
This takes $\cO(|P|)$ time and space.

\begin{figure}[ht]
    \centering
    \includegraphics[width=0.8\linewidth]{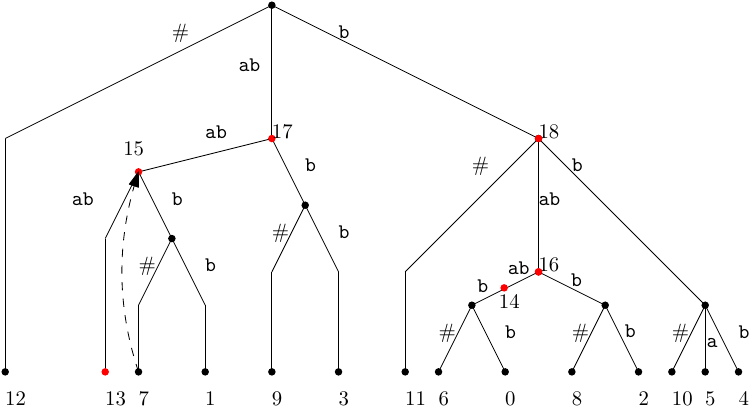}
    \caption{The suffix tree of $T\#P=\texttt{bababbbababb}\#\texttt{ababab}$. The label of edges leading to leaf nodes is truncated to avoid cluttering the figure. The nodes representing suffixes of $P$ are colored red.
    The longest suffix of $P$ occurring as a prefix of $T[p\dd |T|)$ for $p=7$ is the node associated with $15$.}
    \label{fig:ST-del}
\end{figure}

We then employ \Cref{obs:del} (inspect \Cref{fig:ST-del}). We construct the suffix tree $\ST(T\#P)$ of $T\#P$, $\#\notin \Sigma$, and mark the nodes of the suffix tree that represent the suffixes of $P$. The longest suffix of $P$ occurring as a prefix of $T[p\dd |T|)$, for all $p$, is the deepest marked ancestor of the node representing $T[p\dd |T|)$ in $\ST(T\#P)$. Using a bottom-up traversal on the suffix tree and a stack to maintain the currently active marked nodes, we store a pointer from every node $v$ representing a suffix of $T$ to its deepest marked ancestor (i.e., the longest suffix of $P$ represented by an ancestor of node $v$).  By reversing $T$ and $P$, we also store the longest prefix of $P$ that is a suffix of the prefix $T[0\dd p]$ of $T$, for all $p$. This preprocessing takes $\cO(n)$ time and space~\cite{DBLP:journals/jacm/Farach-ColtonFM00}. 

We construct the suffix tree $\textsf{ST}(T)$ and the suffix array $\textsf{SA}(T)$ of $T$. This preprocessing takes $\cO(n)$ time and space~\cite{DBLP:journals/jacm/Farach-ColtonFM00}. We also store at each node of $\textsf{ST}(T)$ the corresponding $\textsf{SA}(T)$ interval -- this can be achieved in $\cO(n)$ time using a bottom-up traversal on $\textsf{ST}(T)$. We also construct a data structure for answering any RMQ on $\textsf{SA}(T)$ and a data structure for answering any RmQ on $\textsf{SA}(T)$. This preprocessing takes $\cO(n)$ time and space~\cite{DBLP:conf/latin/BenderF00}. We also find the interval on $\textsf{SA}(T)$ encoding the occurrences of $P$ in $T$. This takes $\cO(|P|)$ time.

\subsection{Block Deletion}\label{subsec:del}

Again, it is readily verifiable that only Type 1, 2, and 5 occurrences are relevant.
Assume that fragment $T[q\dd p]$ is deleted from $T$.
In particular, we have $T^{i}=T[0\dd q)T[p+1\dd |T|)$. 
In addition, assume that $|T^{i}|\geq |P|$. (If this is not the case, no occurrence of $P$ in $T^{i}$ is to be reported.)
For the occurrences of $P$ in $T[0\dd q)$ (\textbf{Type 1}) we use RmQs and for the occurrences of $P$ in $T[p+1\dd |T|)$  (\textbf{Type 2}) we use RMQs, similar to \Cref{sec:indexing}. 
Namely, each occurrence is reported in $\cO(1)$ time.
For the occurrences of $P$ with a starting position in 
$[0, q)$ and an ending position in $[q, |T^{i}|)$ (\textbf{Type 5}), we retrieve the longest suffix of $P$ that is a prefix of $T[p+1\dd |T|)$ and the longest prefix of $P$ that is a suffix of $T[0\dd q)$, and trigger a prefix-suffix query. The total time to process the deletion is $\cO(1)$ and for reporting is $\cO(\textsf{Occ})$. 
After reporting the occurrences of $P$ in $T^{i}$, we do nothing (to revert the edit).
The correctness follows from \Cref{obs:del}. We have arrived at \Cref{the:pm-del}.

\begin{example}
Let $T=\texttt{b\underline{abab}bb\underline{abab}b}$ and $P=\texttt{ababab}$.
Assume that fragment $T[q\dd p]=T[5\dd 6]=\texttt{bb}$ is deleted from $T$. We have $T^{i}=\texttt{b\underline{abababab}b}$.
We do not have an occurrence of $P$ in $T^{i}$ of Type 1 or 2.
For Type 5 occurrences, we retrieve the longest suffix of $P$ which is a prefix $U$ of $T[p+1\dd |T|)=T[7\dd |T|)$ (inspect \Cref{fig:ST-del}). This is $\texttt{abab}$ (underlined in $T$). We also retrieve the longest prefix of $P$ which is a suffix $V$ of $T[0\dd q)=T[0\dd 4]$, which is also \texttt{abab} (underlined in $T$). We then trigger a prefix-suffix query, which reports two occurrences of $P$ in $UV=\texttt{abab}\cdot \texttt{abab}$ (underlined in $T^{i}$). Indeed, $P$ occurs two times in $T^{i}$.
\end{example}

\section{Pattern Matching with Ephemeral Substring Edits}\label{sec:app-pm-edit}

In this section, we prove \Cref{the:pm-edit}.
We design a data structure for pattern matching with ephemeral substring insertions, deletions, or substitutions. First, we preprocess two strings $T$ and $P$ each of length at most $n$,  over alphabet $\Sigma=[0,\sigma)$ with $\sigma=n^{\cO(1)}$. Then, we allow any ephemeral sequence of edit operations in $T$. Before reverting the $i$th operation, we report all \textsf{Occ} occurrences of $P$ in $T^i$ in the optimal $\cO(\textsf{Occ})$ worst-case time.
This improves upon \Cref{the:dynamic} for this setting.

In \Cref{subsec:edit-preprocess}, we show the preprocessing of $T$ and $P$. In \Cref{subsec:pm-edit}, we show the algorithm for processing the edit operations.

\subsection{Preprocessing}\label{subsec:edit-preprocess}

We start by constructing the data structure of \Cref{the:PSQ} for prefix-suffix queries over $P$. This takes $\cO(|P|)$ time and space. We also construct the String Matching Automaton (SMA) of the \emph{reverse} of $P$. The SMA of a string $X$, denoted by $\textsf{SMA}(X)$, is the minimal deterministic automaton that accepts the language $\Sigma^{*}X$, where $X\in\Sigma^*$. It has $|X|+1=\cO(|X|)$ states, one for every prefix of $X$, with $\varepsilon$ being the \emph{initial} state and $X$ being the \emph{accepting}
state. The crucial property of SMA is that only the transitions that do not lead to the initial state are stored. There are at most $2|X|=\cO(|X|)$ of them, regardless of the size of the alphabet~\cite{DBLP:books/daglib/0020103}. Moreover, the $\textsf{SMA}(X)$ can be constructed in $\cO(|X|)$ time~\cite{DBLP:books/daglib/0020103}; its transitions can be accessed in $\cO(1)$ time if stored using perfect hashing~\cite{DBLP:journals/jacm/BenderCFKT23}. This preprocessing takes $\cO(|P|)$ time and space.

\begin{figure}[t]
    \centering
    \includegraphics[width=0.8\linewidth]{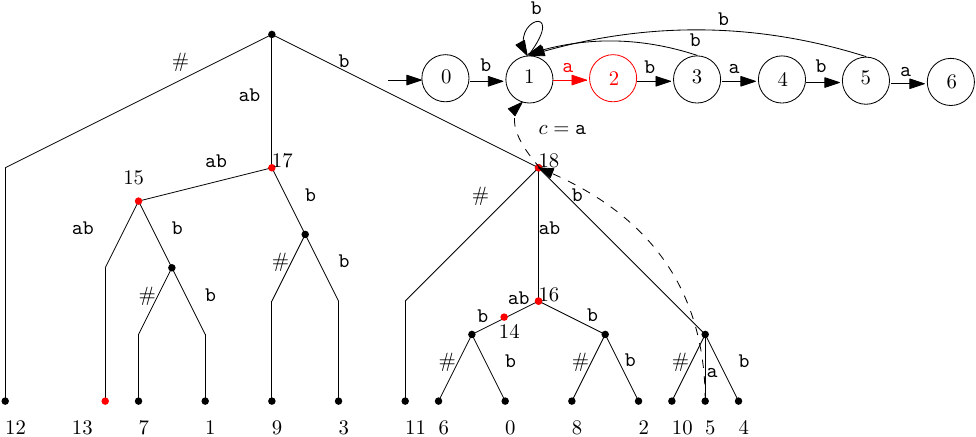}
    \caption{On the left is the suffix tree of $T\#P=\texttt{bababbbababb}\#\texttt{ababab}$. The label of edges leading to leaf nodes is truncated to avoid cluttering the figure. The nodes representing suffixes of $P$ are colored red.
    On the top right is the SMA of the reverse of $P$.
    Consider the query $p=4$ and $c=\texttt{a}$.
    The pointer from the leaf node labeled $p+1=5$ of the suffix tree takes us to its deepest colored ancestor labeled $18$; the corresponding state in the SMA is 1. From there, if we follow the $\texttt{a}$ transition, it takes us forward to state 2, meaning that the longest suffix of $P$ preceded by $c=\texttt{a}$, represented by an ancestor of the leaf node labeled $5$, is $\texttt{b}$, which yields $\texttt{ab}$.}
    \label{fig:ST-ins}
\end{figure}

Deletions are handled by \Cref{the:pm-del}; we thus shift our focus to preprocessing for insertions. We employ \Cref{obs:ins} (inspect \Cref{fig:ST-ins}). We construct the suffix tree $\ST(T\#P)$ of $T\#P$, $\#\notin \Sigma$, and mark the nodes of the suffix tree that represent the suffixes of $P$. 
Every marked node representing $P[i\dd |P|)$ has a pointer to the state of the SMA
representing the reverse of $P[i\dd |P|)$.
Using a bottom-up traversal on the suffix tree and a stack to maintain the currently active marked nodes, we store a pointer from every node $v$ representing a suffix of $T$ to its deepest marked ancestor (i.e., the longest suffix of $P$ represented by an ancestor of node $v$). 
We also store the longest prefix of $P$ which is a suffix of the prefix $T[0\dd p]$ of $T$, for all $p$ (see \Cref{subsec:preprocess}). 
This preprocessing takes $\cO(n)$ time and space. 

We construct the suffix tree $\textsf{ST}(T)$ and the suffix array $\textsf{SA}(T)$ of $T$. This preprocessing takes $\cO(n)$ time and space~\cite{DBLP:journals/jacm/Farach-ColtonFM00}. We also store at each node of $\textsf{ST}(T)$ the corresponding $\textsf{SA}(T)$ interval -- this can be achieved in $\cO(n)$ time using a bottom-up traversal on $\textsf{ST}(T)$. We also construct a data structure for answering any RMQ on $\textsf{SA}(T)$ and a data structure for answering any RmQ on $\textsf{SA}(T)$. This preprocessing takes $\cO(n)$ time and space~\cite{DBLP:conf/latin/BenderF00}. We also find the interval on $\textsf{SA}(T)$ encoding the occurrences of $P$ in $T$. This takes $\cO(|P|)$ time.

\subsection{Edit Operation}\label{subsec:pm-edit}

For simplicity, let us start by edit operations involving a single letter;
it is readily verifiable that only Type 1, 2, 5, and 6 occurrences are relevant.
Assume that we insert letter $c$ after $T[p]$.
If $|P|=1$, a \textbf{Type 6} occurrence is trivially reported in $\cO(1)$ time (otherwise, this type of an occurrence is not relevant).
For the occurrences of $P$ in $T[0\dd p]$ (\textbf{Type 1}) we use RmQs and for the occurrences of $P$ in $T[p+1\dd |T|)$  (\textbf{Type 2}) we use RMQs, similar to \Cref{sec:indexing}. 
Namely, each occurrence is reported in $\cO(1)$ time.
For \textbf{Type 5}, we find the longest suffix of $P$ occurring as a prefix of $cT[p+1\dd |T|)$ as follows. When $p+1$ is given, we find the lex-rank $r=\textsf{iSA}[p+1]$ of $T[p+1\dd |T|)$ among all suffixes of $T$. We find the node $v_{p+1}$ representing $T[p+1\dd |T|)$ in $\ST(T\#P)$ and follow the pointer to its deepest marked ancestor $u_{p+1}$. From there,
using letter $c$, we find the longest suffix $cP[i\dd |P|)$ of $P$ occurring as a prefix of $cT[p+1\dd |T|)$ in $\cO(1)$ time as follows. We follow the pointer of $u_{p+1}$ to the SMA. From the state corresponding to the suffix of $P$, represented by $u_{p+1}$, we follow the transition (if any) using letter $c$. This brings us to the state that represents the reverse of $cP[i\dd |P|)=P[i-1]P[i\dd |P|)$.
Recall that to find the longest prefix of $P$ that is a suffix of $T[0\dd p]$ we do the same work as for the deletion operation on the reverse of $P$ and $T$. After finding the longest prefix of $P$ that is a suffix of $T[0\dd p]$ and the longest suffix of $P$ occurring as a prefix of $cT[p+1\dd n)$, we trigger one prefix-suffix query. The total time to process the insertion is $\cO(1)$ and for reporting is $\cO(\textsf{Occ})$. The algorithm is correct by \Cref{obs:ins}. 

\begin{example}
Let $T=\texttt{bababbbababb}$ and $P=\texttt{ababab}$.
Assume that we insert letter $c=\texttt{a}$ after $T[p]=T[4]=\texttt{b}$. We have $T^{i}=\texttt{b\underline{abab\textcolor{red}{a}b}bababb}$.
We do not have an occurrence of $P$ in $T^{i}$ of Type 1, 2 or 6.
For Type 5 occurrences, we retrieve the node representing $T[5\dd |T|)$ (inspect \Cref{fig:ST-ins}). We follow the pointer to its deepest marked ancestor labeled $18$. Using letter $c=\texttt{a}$, we retrieve $\texttt{ab}$ as the longest suffix of $P$ occurring at position $5$ of $T^{i}$. We also retrieve the longest prefix of $P$ that is a suffix $T[0\dd p]=T[0\dd 4]$, which is \texttt{abab}. We then trigger a prefix-suffix query, which reports an occurrence of $P$ in $T^{i}$ (underlined).
\end{example}

When handling a string $S$ of length $|S|=\cO(1)$ (instead of a single letter $c$), the only difference for \textbf{Type 5} is that from the SMA state corresponding to the suffix of $P$, we follow the transition \emph{path} using the reverse of $S$. This brings us to the state that represents the reverse of $S\cdot P[i\dd |P|)$.
This is correct because, for any strings $U$ and $X$, the 
$\textsf{SMA}(X)$ takes us to the longest prefix of $X$ that is a suffix of $U$. 
Thus, when $X$ is the reverse of $P$,
and $U$ is the reverse of $S\cdot T[p+1\dd |T|)$,
the SMA takes us to the longest suffix of $P$
that is a prefix of $S\cdot T[p+1\dd |T|)$.
\textbf{Type 3}, \textbf{Type 4}, and \textbf{Type 6} occurrences
are handled exactly as in \Cref{sec:indexing}. Finally, recall that substitutions work the same way as insertions, but with a different offset. We have arrived at \Cref{the:pm-edit}.

\section{Concluding Remarks}\label{sec:fin}

We presented algorithms for text indexing and pattern matching with ephemeral edits. Our algorithms are simple to implement as they utilize only textbook algorithms or data structures with mature implementations; and they also outperform their fully dynamic counterparts, while, at the same time, they are founded on a model highly motivated by real-world applications.

It would be interesting to improve \Cref{the:dynamic} by improving the preprocessing cost for the pattern or improving the query time for reporting. Another direction would be to consider other versions of text indexing or pattern matching with ephemeral edits (here we have considered their vanilla version).

\section*{Acknowledgments}
This work was supported by the PANGAIA and ALPACA projects that have received funding from the European Union’s Horizon 2020 research and innovation programme under the Marie Skłodowska-Curie grant agreements No 872539 and 956229, respectively.

\bibliographystyle{alpha}
\bibliography{references}

\end{document}